\documentclass[useAMS,usenatbib, usegraphicx]{mn2e}
\usepackage{epsfig}
\usepackage{fancyhdr}
\usepackage{makeidx}
\usepackage{titletoc}
\usepackage{lscape}
\usepackage{hyperref}
\usepackage{wasysym}
\usepackage{wrapfig}
\usepackage{longtable}
\usepackage{aas_macros}
\usepackage{txfonts}
\usepackage{times}
\begin{document}

\title[The star formation history of the Sagittarius stream]{The star formation history of the Sagittarius stream}
\author[T.J.L. de Boer, V. Belokurov and S. Koposov]{T.J.L. de Boer$^{1}$\thanks{E-mail:
tdeboer@ast.cam.ac.uk}, V. Belokurov$^{1}$ and S. Koposov$^{1}$\\
$^{1}$ Institute of Astronomy, University of Cambridge, Madingley Road, Cambridge CB3 0HA, UK}
\date{Received ...; accepted ...}

\pagerange{\pageref{firstpage}--\pageref{lastpage}} \pubyear{2014}

\maketitle

\begin{abstract}
We present the first detailed quantitative study of the stellar
populations of the Sagittarius~(Sgr) streams within the Stripe 82
region, using photometric and spectroscopic observations from the
Sloan Digital Sky Survey~(SDSS). The star formation history (SFH) is determined separately for the bright and faint
Sgr streams, to establish whether both components consist of a similar
stellar population mix or have a distinct origin.  

Best fit SFH solutions are characterised by a
well-defined, tight sequence in age-metallicity space, indicating
that star formation occurred within a well-mixed, homogeneously enriched medium. 
Star formation rates dropped sharply at an age of $\approx$5-7
Gyr, possibly related to the accretion of Sgr by the MW. 
Finally, the Sgr sequence displays a
change of slope in age-metallicity space at an age between 11-13 Gyr
consistent with the Sgr $\alpha$-element knee,
indicating that supernovae type Ia started contributing to the
abundance pattern $\approx$1-3 Gyr after the start of star formation.

Results for both streams are consistent with
being drawn from the parent Sgr population mix, but at
different epochs. The SFH of the bright stream starts from old,
metal-poor populations and extends to a metallicity of
[Fe/H]$\approx$$-$0.7, with peaks at $\approx$7 and 11 Gyr. The
faint SFH samples the older, more metal-poor part of the Sgr
sequence, with a peak at ancient ages and stars mostly with
[Fe/H]$<$$-$1.3 and age$>$9 Gyr. Therefore, we argue in favour of a
scenario where the faint stream consists of material stripped i)
earlier, and ii) from the outskirts of the Sgr dwarf.
\end{abstract}


\label{firstpage}

\section{Introduction}
\label{introduction}

The promise of Galactic Archaeology is that the physical conditions in
the early Universe, at high redshifts corresponding to the initial
bouts of the Milky Way (MW) formation can be gleaned from studies of
nearby stars in unrelaxed fragments of its tidally shredded satellites
\citep[see e.g.][]{Freeman02}. In principle, if enough pieces of a
destroyed dwarf galaxy can be identified in the Milky Way halo and put
back together, its star-formation and chemical enrichment histories
can all be reconstructed. Not only could this help to shed light onto
the nature of the abundance differences between the destroyed and the
surviving MW satellites \citep[see e.g.][]{Tolstoy09}, but also to
understand the punctuation of the Galactic accretion history. The
latter is of course impossible without an idea of how and when the
satellite destruction occurred. Luckily, changes in chemistry and
star-forming activity can be matched to the structural properties of
the tidal debris to break degeneracies associated with the satellite
disruption modelling.

Even though the concept of such archaeological measurement has been the
point of discussion for some time, in practice, no attempt has been
made so far to infer star-formation and metal-enrichment histories of
a disrupting (or fully undone) satellite system. This is simply due to
the low-surface brightness nature of the Galactic stellar halo
sub-structure accompanied by high levels of contamination from the
disk and the rest of the halo. In this paper we turn to the
Sagittarius (Sgr) stream, the largest known sub-structure in the Milky
Way halo \citep[see e.g.][]{Belokurov13}. According to the most
recent studies \citep[][]{Deason11,Deason14}, the Sgr stream can
easily contribute $\sim10\%$ of the total luminosity of the stellar
halo and it appears to dominate the faint Main Sequence Turn-off
(MSTO) star counts in the regions of the sky falling within its
orbital plane. Therefore, we take advantage of the large number of
photometric and spectroscopic Sgr stream tracers available to carry
out a first study of the assembly history of a stellar system
dispersed by tides.

The progenitor of the stream, the Sgr dSph galaxy was discovered
by~\citet{Ibata95} and to the present day holds the record of the
nearest dwarf galaxy to the Milky Way~(MW), at a distance
of~$\approx$15 kpc from the Galactic centre. With an estimated total
luminosity of $\approx$10$^8$ L$_{\odot}$ and its dark matter~(DM) mass of
$\approx$10$^{9}$ M$_{\odot}$ before infall, the Sgr dSph is the third
most luminous and the third most massive recognisable satellite of the
MW, after the Large and Small Magellanic
clouds~\citep{Niederste-Ostholt10}. An array of studies spanning now
almost two decades, showed that the system is currently being accreted
by the MW, undergoing severe stripping due to the effects of Galactic
tides~\citep[see a compendium of references in][]{Koposov15}. It is
now certain that a large fraction of the stars belonging to Sgr have
been stripped from the outskirts and the core to form large stellar
streams wrapping around the Galaxy at least once. Note that according
to the recent study of \citet{Belokurov14}, the trailing tail extends
further than originally thought, increasing the total arc covered by
the debris beyond $2\pi$.

To date, several different Sgr stream models have been presented
\citep[e.g.,][]{Fellhauer06, MartinezDelgado07, Law10a,
  Penarrubia10,Gibbons14}. With various degrees of success, these
explain some, but never all, properties of the stream. What has
muddled the picture of the dwarf disruption is the co-existence of the
Virgo overdensity ~\citep[e.g.][]{Juric08} and a significant portion
of the leading tail around the North Galactic cap. Additionally, both
leading and trailing tail appear to be bifurcated into two distinct
components ~\citep{Belokurov06,Koposov12}. The results of the 3D
tomography of the arc of the leading tail in the North presented by
\citet{Belokurov06} and later verified by \citet{Yanny091} seem to
rule out the connection between the Sgr stream and the Virgo
Cloud. However, the nature of the bifurcation of the Sgr tails remains
concealed.

Attempts have been made to identify clear differences in the
properties of the bright and the faint components of the bifurcation
\citep[see e.g.][]{Belokurov06, Yanny091, Niederste-Ostholt10,
  Koposov12, Slater13}.  All studies agree that the distances and the
line-of-sight velocity match nearly perfectly, albeit with small but
noticeable deviations.  With regards to chemistry, on the other hand,
there is a good deal of dispute.  For example, according to
\citet{Yanny091}, there is no obvious disagreement between the
stellar population compositions of the bright and the faint components
of the leading tail bifurcation.  This is contested by
\citet{Koposov12}, who with the help of photometric metallicities argue
for the presence and the lack of a strong metal-rich component in the
bright and faint streams correspondingly, in both leading (in the
North) and trailing (in the South) tails.  Note that, the only
plausible interpretation of the bifurcation which has not been
completely ruled out invokes a rotating, disky Sgr progenitor
\citep{Penarrubia10}.  Any metallicity differences between portions of
the stream can therefore be easily compared to the models of
disrupting disks with realistic $[$Fe/H$]$ gradients.

Given its large estimated total mass prior to disruption, Sgr is an
excellent laboratory for studies of star formation in systems at the
upper end of the MW dwarf galaxy mass distribution.  Accordingly,
metallicity across the face of the remnant and along the streams have
been scrutinised. For example, ~\citet{Bellazzini062}, based on the
distribution of stars on the horizontal branch within the main body
and the stellar streams determine that Sgr displays a population
gradient as a function of radius, similar to what has been observed in
other LG dwarf galaxies. Studies of individual RGB stars within the
stream and dwarf galaxy have revealed that Sgr is a metal-rich system
with stars reaching close to solar metallicities in the very
centre~\citep[e.g.,][]{Monaco05, Sbordone07, Carretta102,
  McWilliam13}. The spectroscopic $\alpha$-element distribution is
characterised by a plateau at low metallicities followed by a distinct
$\alpha$-element ''knee" at [Fe/H]=$-$1.3, consistent with chemical
enrichment in a massive dwarf galaxy~\citep{deBoer2014A}.

Studies of the full colour magnitude diagram~(CMD) of Sgr in the main
body have been used to place limits on the age and metallicity of its
stellar populations using comparisons with stellar tracks and
isochrones~\citep{Bellazzini06,Bellazzini08}. However, no detailed
modelling of Sgr CMDs has yet been done, to determine quantitative
star formation rates of individual stellar populations. It is only a
complete modelling of the distribution of stellar populations present
in the CMD that can determine the detailed formation history of Sgr,
including the presence of distinct star formation peaks and the width
of the stellar population locus in age and metallicity space as
induced by inhomogeneous mixing.
\begin{figure*}
\centering
\includegraphics[angle=0, width=0.49\textwidth]{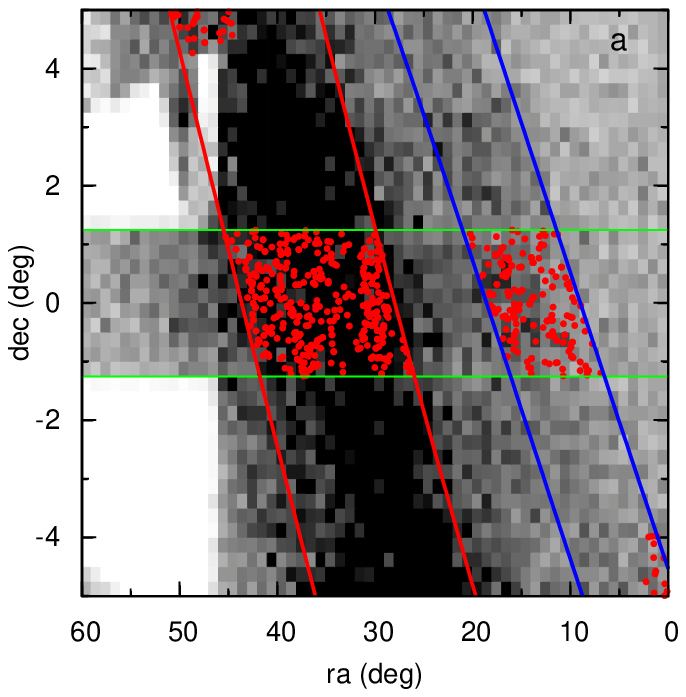}
\includegraphics[angle=0, width=0.49\textwidth]{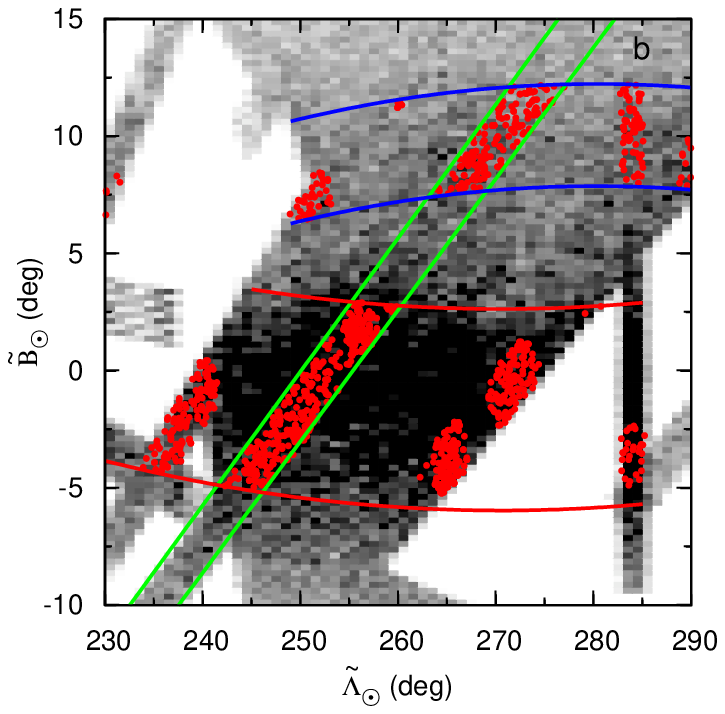}
\caption{Spatial coverage of the photometric and spectroscopic observations across the Sgr stream around SDSS Stripe 82, in ra,dec coordinates~(\textbf{a}) and in coordinates aligned with the Sgr stream~(\textbf{b}). The greyscale density plots show the density of stars~(0.2$\le$g$-$i$\le$0.45, 20.0$\le$i$\le$21.5) while the red points show the coverage of spectroscopic metallicities. Green lines indicate the extent of SDSS Stripe 82, while red and blue lines indicate the selection we adopt for the bright and faint stream respectively. \label{Sgrspatial}}
\end{figure*}

This is indeed the focus of our work: to study the detailed star
formation history (SFH) of the Sgr stream in the Stripe~82 region,
using data from the SDSS Data Release 10~\citep{Ahn14} and the Sloan
Extension for Galactic Understanding and Exploration~(SEGUE). Deep, MW
foreground corrected CMDs will be used to determine the stellar
population make-up of the stream in comparison to the populations of
the parent Sgr galaxy. We utilise deep co-added photometry of the
extensively covered Stripe~82 region to determine the photometric
completeness of the single-epoch SDSS catalogs, allowing us to
accurately model the observed CMDs~\citep{Annis11}. Furthermore,
spectroscopic metallicities of individual stars are directly used in
combination with the photometry to break the age-metallicity
degeneracy and provide additional constraints on the age of the
stellar populations, as described in~\citet{deBoer2012A}. We
determine the detailed SFH for the bright and faint streams
separately, to determine if both stream components consist of the same
population mix of if they constitute two different streams.

The paper is structured as follows: in section~\ref{photometry} we
present the selection of Sgr stars from the photometric SDSS survey,
the MW foreground correction applied and the photometric completeness
determination. The determination of the spectroscopic metallicity
distribution of the Sgr streams is described in
Section~\ref{spectroscopy}. Section~\ref{SFHmethod} describes the
determination of the SFH and discusses the specifics of fitting Sgr
CMDs. The detailed SFH of the bright Sgr stream is presented in
Section~\ref{brightstream}, followed by the SFH determination of the
faint stream in Section~\ref{faintstream}. Finally,
Sect.~\ref{conclusions} discusses the results obtained from the
comparison of both stream components and the implications from the SFH
for the formation history of the Sgr dSph.
\begin{figure}
\centering
\includegraphics[angle=0, width=0.495\textwidth]{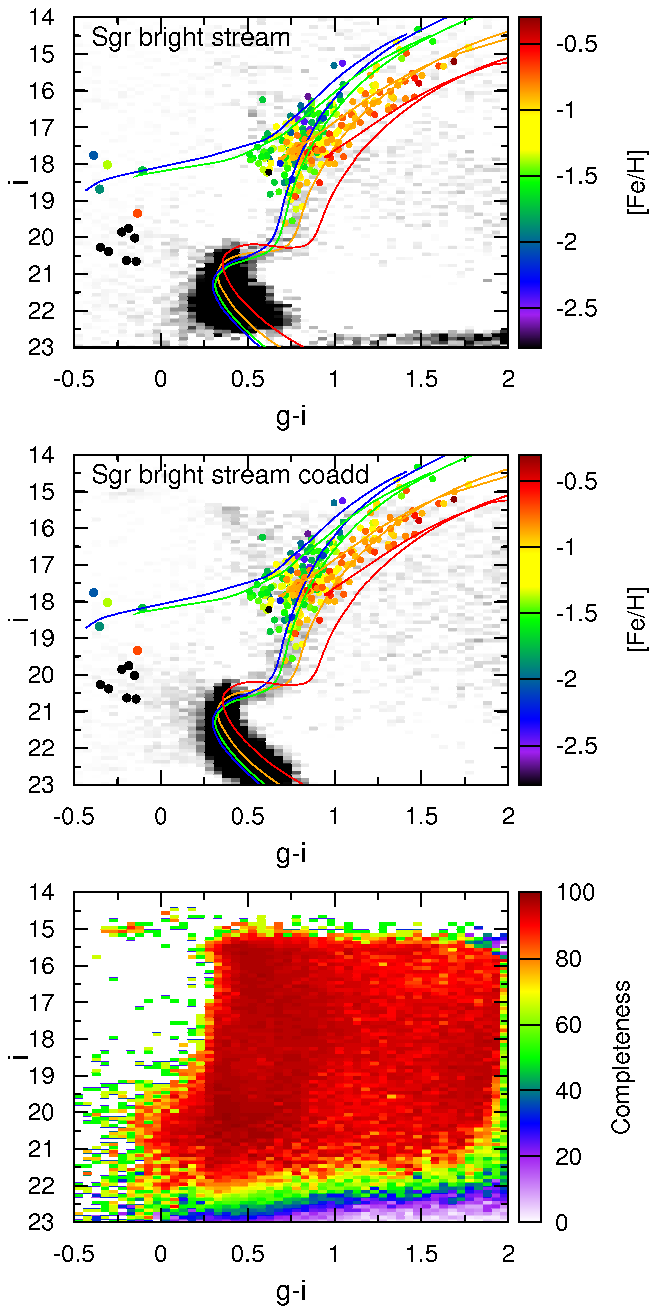}
\caption{Comparison between extinction corrected single-epoch SDSS photometry~(upper panel) and co-added photometry~(middle panel) for the bright Sgr stream. Greyscale Hess diagrams show the density of stars in the CMD while the coloured points show spectroscopic observations of stream stars with colour indicating the metallicity. Several reference isochrones and HB tracks are also shown, corresponding to metal-poor~(blue, [Fe/H]=-2.0, age=13 Gyr), intermediate~(green, [Fe/H]=-1.5, age=12 Gyr) and metal-rich populations~(orange: [Fe/H]=-0.8, age=8 Gyr, red: [Fe/H]=-0.3, age=5 Gyr) within Sgr. The photometric completeness is shown in the bottom panels, calculated by comparing the single-epoch and co-added data. \label{SgrbrightCMD}}
\end{figure}

\section{Sgr photometric data}
\label{photometry}
To obtain accurate photometry of the Sgr stream in the Southern Stripe~82 we make use of SDSS Data Release 10~\citep{Ahn14}. SDSS covered a wide area of the Sgr stream in the Southern hemisphere, with a 275 deg$^{2}$ degree region (equatorial Stripe~82) in particular being covered during numerous epochs. This region was covered during multiple runs resulting in deep co-added catalogs of stars reaching $\approx$2 magnitudes fainter than the SDSS single pass data~\citep{Annis11}. This extra depth is crucial for SFH analysis, as we can assume that the co-added catalog constitutes a 100\% complete sample at the faintest magnitudes probed by single pass data. Therefore, a comparison between both catalogs can be used to estimate the photometric completeness in SDSS Stripe~82 data, a critical component of synthetic CMD analysis. 
\begin{figure}
\centering
\includegraphics[angle=0, width=0.495\textwidth]{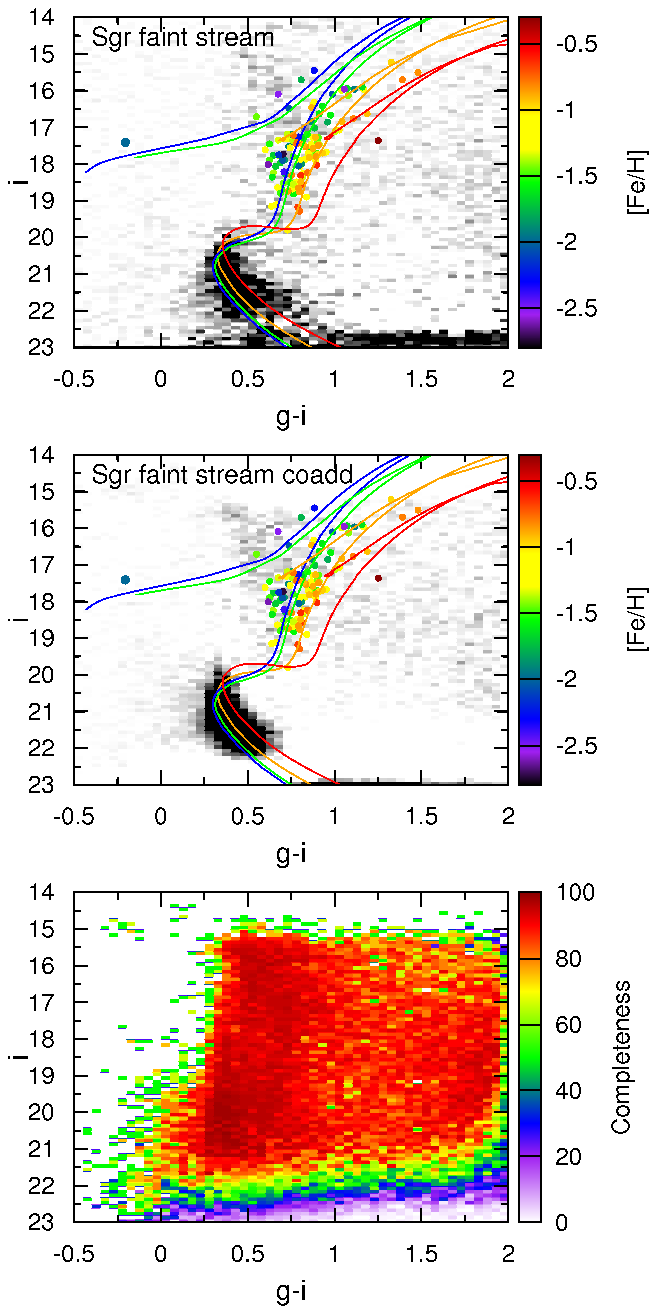}
\caption{Same as Figure~\ref{SgrbrightCMD} but for the faint Sgr stream. \label{SgrfaintCMD}}
\end{figure}

We select our Sgr photometric sample by making use of the
prescriptions of~\citet{Majewski03} to transform equatorial ra and dec
coordinates into heliocentric $\widetilde{\Lambda}_{\odot}$,$\widetilde{\mathrm{B}}_{\odot}$
coordinates that are aligned with the Sgr stream. Following the
nomenclature of~\citet{Belokurov14}, $\widetilde{\Lambda}_{\odot}$ increases in
the direction of Sgr motion and $\widetilde{\mathrm{B}}_{\odot}$ points to the North
Galactic pole. Figure~\ref{Sgrspatial} shows the density of main
sequence~(MS) stars in a region around Stripe~82 in ra, dec
coordinates and coordinates aligned with the Sgr stream. Over-densities
corresponding to the bright and faint Sgr streams are visible in
Figure~\ref{Sgrspatial}, highlighted by red and blue lines respectively. 
We combine these spatial cuts with the extent of Stripe~82~(green lines in Figure~\ref{Sgrspatial})
to select our photometric sample of the Sgr bright and faint streams.

Due to the close proximity of the stream, there might be overlap between the bright and faint streams. To determine the effect of
this overlap, we calculate the fraction of stars from one stream
contained within the spatial selection of the other stream, under the
assumption of Gaussian density distributions. We derive a
contamination fraction of the bright stream within the faint stream
region of 4\%, which is large enough to be noticeable in the CMD. On
the other hand, the contamination fraction of the faint stream on the
bright stream is negligible at 0.05\%. Following these calculations,
we correct for the contamination effect on the faint stream, by
subtracting the weighted CMD of the bright stream.

The CMDs determined in this way will contain stars of the Sgr stream
as well as stars belonging to the MW at a wide range of distances and
colours. These stars will contaminate the Sgr CMD and will be
incorrectly fitted during the SFH determination using models at the
distance of Sgr, leading to anomalously old and/or metal-rich
populations in the SFH solution. To correct for the presence of these
contaminants, we correct the photometric CMDs by subtracting a
representative comparison CMD containing mostly MW stars. To select an
appropriate correction region we consider that the MW is, to first
order, symmetric around the Galactic plane and select stars from SDSS
with the same Galactic longitude but latitude mirrored with respect to
the Galactic plane. In this way we obtain a correction region
containing the appropriate mix of disk and halo stars present at the
Galactic longitude and latitude of our Sgr fields. 

For the deep co-added photometry, we are constrained solely to data from Stripe~82
and therefore we cannot use a mirrored patch of sky in the foreground
correction. Instead, we select a region between $-$10$<$ra$<$0 degrees
within Stripe~82 for our decontamination field. This field corresponds
to similar galactic latitude as our stream fields~($\delta$b$\le$5
deg) but significantly different galactic longitude~($\delta$l$\ge$75
deg for the bright stream). Therefore, the stellar population mix of
the MW foreground region is likely not the same as in our stream
regions. To obtain an optimal subtraction of MW foreground halo
stars~(displaying similar colours as stream MSTO stars), we apply a
colour shift to the foreground CMD before subtracting it from the
stream CMD. The colour shift is derived by comparing the median g$-$i
colour of MW halo turnoff stars in both CMDs using a selection box
with 0.1$<$g$-$i$<$0.5, 18$<$i$<$19.5. 

The Stripe~82 samples of Sgr stars span a total range of $\approx$20 degrees of angle along the
stream. The stream displays a distance gradient of $\approx$5 kpc
across this range, corresponding to a sizeable distance modulus
difference of 0.35 mag. To correct for this variation of distance
along the stream, we determine the distance of the bright and faint
stream based on CMD distance indicators such as BHB, red clump~(RC)
and sub-giant branch~(SGB) stars. Distances determined for both streams
agree within the uncertainties with literature distance determinations
by~\citet{Niederste-Ostholt10,Koposov12}. We correct for the presence
of the distance gradient along both streams by interpolating the
distance to each star in our sample and shifting it up or down in the
CMD to a reference distance modulus of 17.5 mag~(31.6 kpc) for the
bright stream and 17.0 mag~(25.1 kpc) for the faint stream,
corresponding roughly to the observed distance to the stream within
Stripe~82. Finally, to correct for the effect of dust extinction we
also used extinction maps of~\citet{Schlegel98} to determine the
extinction toward each individual star within our sample and created
extinction-free CMDs. The same distance and extinction correction
procedure is also applied to the MW correction sample before
subtracting it from the stream samples.
\begin{figure}
\centering
\includegraphics[angle=0, width=0.495\textwidth]{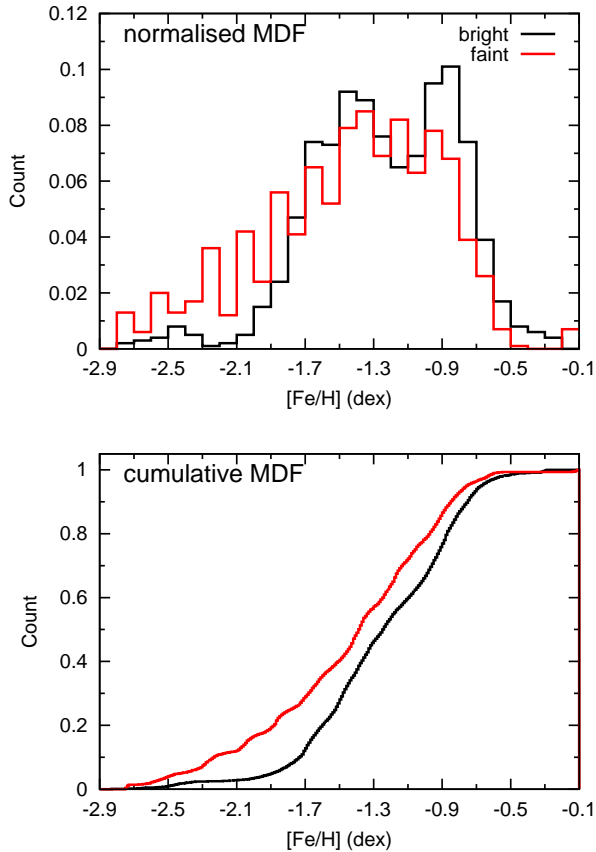}
\caption{The metallicity distribution function of the bright and faint stream as obtained from spectroscopic observations. The top panel shows the normalised MDF, while the bottom panel shows the MDF in cumulative form. \label{MDFobs}}
\end{figure}

The resulting MW corrected CMD of the bright stream is displayed in
Figure~\ref{SgrbrightCMD}. The deep co-added CMD for each region is
also shown as well as the photometric completeness resulting from the
comparison of both catalogs. Additionally, Figure~\ref{SgrfaintCMD}
shows the same CMDs and completeness for the faint Sgr stream. The Sgr
MS, SGB and lower RGB are clearly visible and traced by model
isochrones at the distance of Sgr.

\section{Sgr spectroscopic data}
\label{spectroscopy}

To obtain our Sgr spectroscopic sample, we make use of the SEGUE
survey, which obtained medium-resolution spectra of stars covering a
large portion of the sky, including many fields consistent with the
Sgr stream~\citep{Yanny091}. The [Fe/H] and [$\alpha$/Fe] ratios for
individual stars have been determined through detailed synthetic
spectrum fitting using the SEGUE Stellar Parameter
Pipeline~\citep[SSPP,][]{SEGUE1,SEGUE2,SEGUE3,SEGUE4,SEGUE5}. To
select our sample of Sgr stars, we initially select stars based on
their position on the sky, in the same way as the photometric
sample. Additionally, we limit our sample to stars with robust stellar
parameters and abundances using S/N$\ge$50 as well as
spectroscopically confirmed giants using log $g$$\le$3.5,
4300$\le$T$_{\rm eff}$$\le$6000 K.

To reduce contamination from MW stars in the spectroscopic sample, we
exploit the fact that the Sgr stream displays a clear signal in the
$\widetilde{\Lambda}_{\odot}$ vs line-of-sight velocities V$_{GSR}$
plane~\citep{Belokurov14}. In this plane, the Sgr-stream stars form a
narrow sequence in each hemisphere, with velocities distinct from the
MW. We use these sequences to refine our sample by selecting only
stars consistent with the Sgr signal. To further decontaminate our
sample we determine the distances to individual stars. Distances are
obtained by selecting isochrone points with the correct g$-$i colour,
and shifting their i-band magnitude to match the observations. We make
use of the age-metallicity relation of Sgr globular clusters to
determine the correct age at each
metallicity~\citep{Forbes10}. Subsequently, we pare our sample by
keeping only those stars with distances consistent~(to within 10 kpc) with the literature
distance determinations to different branches of the stream from
different distance
indicators~\citep{Niederste-Ostholt10,Koposov12}. Finally, we shift
the spectroscopic stars of this sample according their distance to the
same reference distance for the bright and faint stream as adopted for
the photometric samples.

The sample of stars obtained in this way are displayed in
Figure~\ref{Sgrspatial} as red points. In this work, we focus on SDSS
Stripe~82 and therefore we only select those stars within the green
lines in Figure~\ref{Sgrspatial}. The stars in the spectroscopic
sample are overlaid on the CMD in
Figures~\ref{SgrbrightCMD} and~\ref{SgrfaintCMD} with colours
representing the spectroscopic [Fe/H] abundance. The stars from our
spectroscopic sample form a clear RGB as a continuation of the MS, SGB
and RGB branches seen in the photometric CMD. Furthermore, a clear
metallicity trend is visible in the CMD, consistent with the evolution
of stellar populations in a relatively isolated system, indicating
that our sample indeed consists of likely Sgr members.

\begin{figure*}
\centering
\includegraphics[angle=0, width=0.9\textwidth]{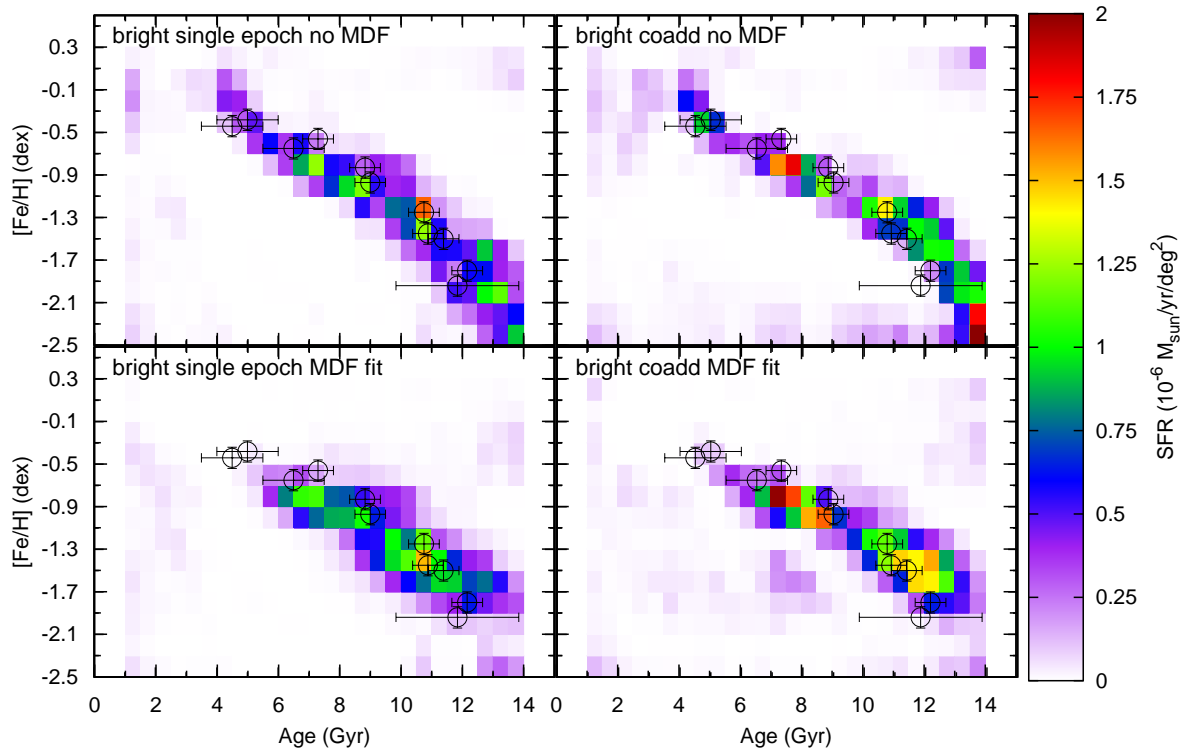}
\caption{Full SFH solution as a function of age and metallicity for the bright Sgr stream within Stripe 82. Left panels show the solution derived from single-epoch photometric data, with and without MDF fitting, while the right panels shows solution derived from deep co-added data. The age and metallicity of globular clusters associated to the Sgr stream are also shown, as determined from deep CMD studies~\citep[][and references therein]{Forbes10}. \label{SFRbright}}
\end{figure*}

To obtain a representative metallicity distribution function~(MDF) for
each stream, we must also take into account the spectroscopic sampling
efficiency of SDSS/SEGUE, which consists of different surveys with
different depths and target selections. Therefore, we correct for the
completeness of spectroscopic stars as a function of position in the
CMD. For each SDSS plate we take all stars~(including MW foreground) with SDSS photometry and all stars with spectroscopy, by sampling the area of each target plate~(with radius
1.49 degrees). Those two samples are binned in the CMD space and the ratio of the two CMD densities tells us the completeness of the SDSS spectroscopy relative to the photometry
, which is assumed to be complete at the sampled
magnitudes. 

These completeness measures are then used to weight the
importance of each star in the observed MDF for the bright and faint
streams. Finally, we also correct the spectroscopic MDF of the faint
stream for the effect of contamination from the bright stream, by
subtracting a weighted bright stream MDF as outlined in
Section~\ref{photometry}. The spectroscopic MDFs obtained in this way are 
shown in Figure~\ref{MDFobs}, highlighting the differences between both streams. 
The cumulative MDF clearly shows that the faint stream is composed of more metal-poor
populations than the bright stream, with almost no stars more metal-poor than [Fe/H]=$-$0.9.

\section{SFH method}
\label{SFHmethod}

The SFH of the Sgr streams will be determined using the routine Talos,
which employs a synthetic CMD
method~\citep[e.g.,][]{Tosi91,Tolstoy96,Gallart962,Dolphin97,Aparicio97}. This
technique determines the SFH by comparing the observed CMDs to a grid
of synthetic CMDs using Hess diagrams~(plots of the density of
observed stars), taking into account photometric error and
completeness. Uniquely, Talos simultaneously takes into account the
photometric CMD as well as the spectroscopic MDF, providing direct
constraints on the metallicity of stellar populations to obtain a
well-constrained SFH. We refer to~\citet{deBoer2012A} for a detailed
description of the routine and performance tests. 

To determine the SFH of Sgr, we assume a wide range of possible ages
and metallicities to avoid biasing the solution by our choice of
parameter space. For the metallicity, a lower limit of [Fe/H]=$-$2.5
dex is assumed, which is the lowest available in the Dartmouth library
used in this work~\citep{DartmouthI}. We do not expect this to lead to
a bias in the SFH results since very few stars in Sgr are expected to
have low metallicities. An upper limit of [Fe/H]=0.3 dex is assumed
for the metallicity based on spectroscopic measurements of the
metallicity in central body of
Sgr~\citep{SmeckherHane02,Carretta102,McWilliam13}. The $\alpha$-element abundance of the isochrones used is varied with metallicity according to the observed [$\alpha$/Fe] distribution of the Sgr stream, ranging from [$\alpha$/Fe]=$-$0.2 to [$\alpha$/Fe]=0.4~\citep{deBoer2014A}. For the age limits, we assume a maximum age of 14 Gyr for the age of the Universe
and considered a range of ages between 0.25~(the youngest age
available in the isochrone set used) and 14 Gyr old, with a uniform
bin size of 0.5 Gyr. 

The SFH is determined using the (i, g$-$i) CMD to make use of a large
dynamic range in the CMD while sticking to the well studied g and
i bands. The SFH fitting is restricted to the MS and lower RGB
regions, using CMD limits of $-$0.3$<$g$-$$i<$1.3 and 18$<$i$<$22 for
the bright stream and $-$0.3$<$g$-$$i<$1.3 and 17.5$<$i$<$21.5 for the
faint stream. The bright magnitude cut excludes the heavily
contaminated upper RGB and MW thick disk regions, while the faint end
of the CMD is truncated to avoid fitting the SFH in regions where
the photometry becomes less reliable in SDSS data~(see
Figure~\ref{SgrbrightCMD}).

\begin{figure*}
\centering
\includegraphics[angle=0, width=0.9\textwidth]{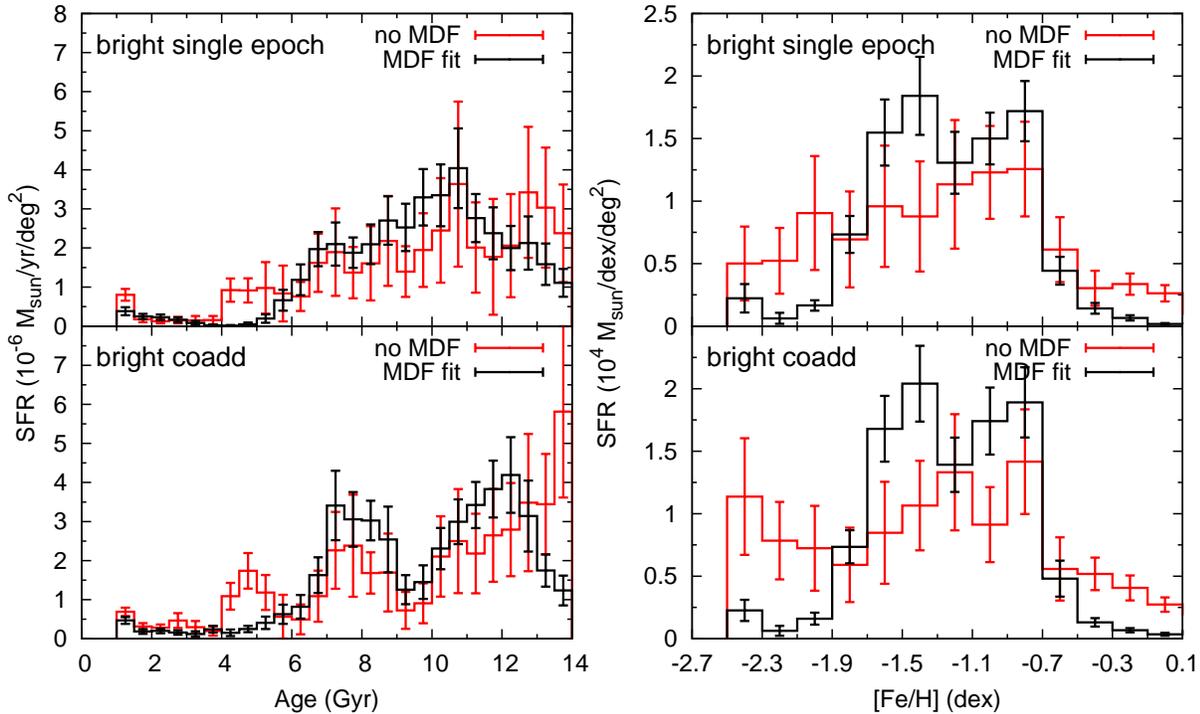}
\caption{Star formation rates projected onto age~(star formation history, left) and metallicity~(chemical evolution history, right) of the bright Sgr stream. Solutions from single-epoch data are shown in the top panels, while co-add photometry solutions are shown in the bottom panels.  \label{SFHbright}}
\end{figure*}

\section{Star formation history of the bright stream}
\label{brightstream}

The star formation history of the Sgr bright stream is derived using
the single-epoch SDSS data presented in Figure~\ref{SgrbrightCMD}. The
SFH solution is obtained both with and without fitting the
spectroscopic MDF, to assess the effect of the MDF constraints on the
derived stellar populations. Furthermore, to study the effects of data
quality and depth on the SFH, we also make direct use of the deep
co-added photometry available in SDSS Stripe 82. Therefore, we also
determine the SFH from the co-add photometry for the bright stream,
under the assumption that the photometry is 100\% complete down to the
faint CMD limit of i=22.

Figure~\ref{SFRbright} shows the SFH solutions as obtained from both
the single-epoch~(left panels) and deep co-added SDSS
photometry~(right panels) for the Sgr bright stream. Solutions
obtained with and without fitting the MDF are shown in the bottom and
top panels respectively, to highlight the differences in the SFH
associated to the constraints from the spectroscopic MDF. The star
formation rate~(SFR) is normalised by spatial extent using a total
area of 39.4 deg$^{2}$ for the bright stream as shown in
Figure~\ref{Sgrspatial}. 

\begin{figure*}
\centering
\includegraphics[angle=0, width=0.45\textwidth]{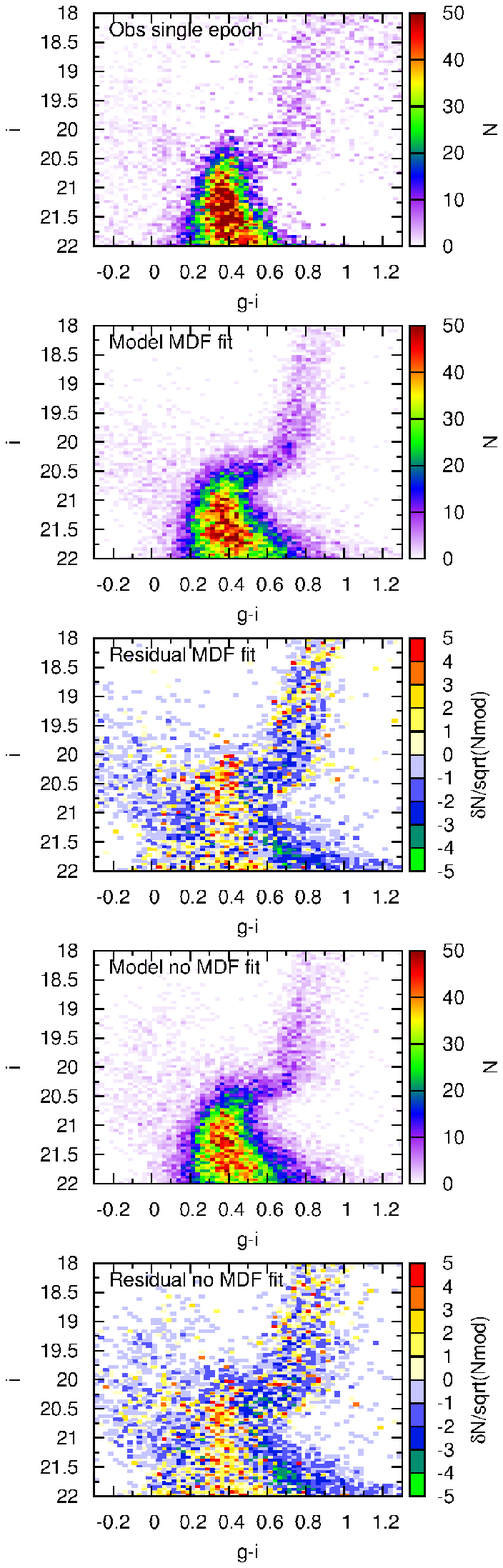}
\includegraphics[angle=0, width=0.45\textwidth]{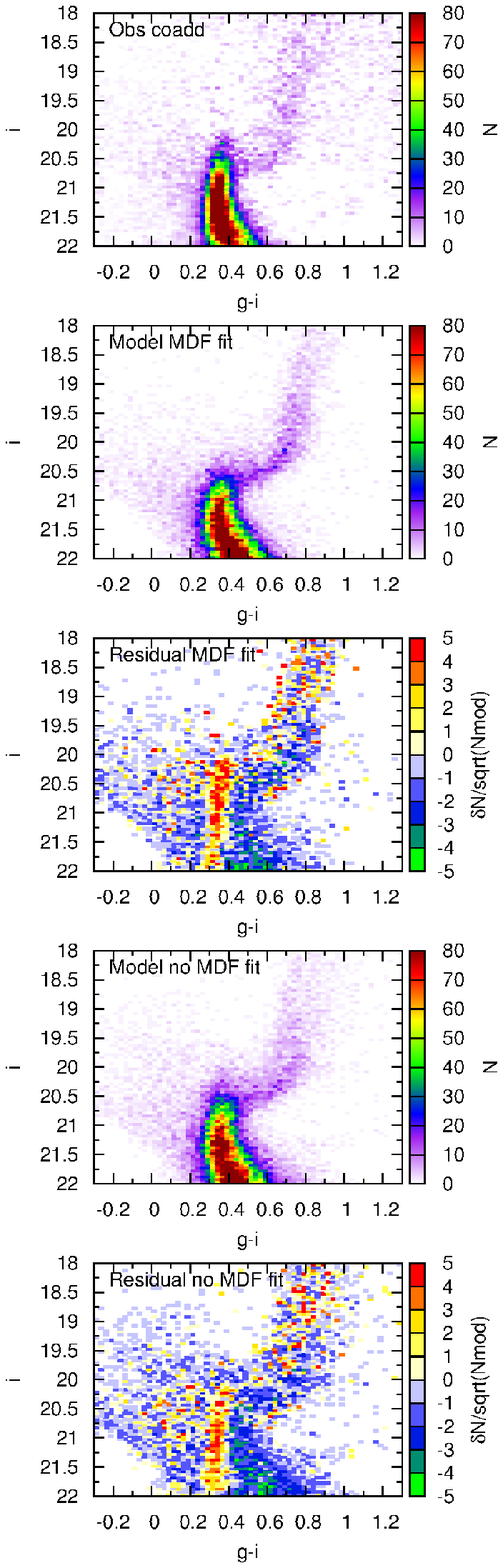}
\caption{Observed and best-fitting CMDs for the bright stream of Sgr, with and without MDF fitting. The middle and bottom rows show the difference between the observed and best-fit CMDs, expressed as a function of the uncertainty in each CMD bin. \label{Hessbright}}
\end{figure*}

By projecting the SFR values onto one axis we obtain the SFR as a
function of age~(SFH) or metallicity~(chemical evolution history,
CEH), which are shown in Figure~\ref{SFHbright}. The error bars
indicate the uncertainty on the SFR as a result of different CMD and
parameter grids~(as described in~\citet{deBoer2012A}). The SFH and CEH
display the rate of star formation at different ages and metallicities
over the range of each displayed bin in units of solar mass per year
or dex respectively. The total stellar mass in this section of the
stream can be computed by multiplying the star formation rates by the
range in age or metallicity of each bin and multiplying by the total
spatial area.

To assess the goodness of fit of the SFH solutions,
Figure~\ref{Hessbright} shows a comparison between the observed and
synthetic (i, g$-$i) CMDs of the bright stream for the single-epoch
data~(left panels) and the deep co-added data~(right panels). The
middle and bottom panels of Figure~\ref{Hessbright} show the fit
residuals in each bin in terms of Poisson
uncertainties. Furthermore, the goodness of fit can also be assessed
by comparing the metallicities of stars in the best-fit SFH model on the upper RGB to the observed MDF
as shown in Figure~\ref{MDFbright} with and without MDF fitting.


\subsection{Single-epoch data}
\label{brightsingleepoch}
The SFH solutions derived from single-epoch photometry allow us to
study the Sgr stream populations using data of a depth and quality as
found throughout the SDSS survey. In this way, we can study the
formation history of Sgr with a precision which can also be achieved
at different angles along the stream within SDSS.

To assess the goodness of the derived fit, we compare the observed and synthetic CMDs of the bright stream in the left panels of Figure~\ref{Hessbright}. Comparison between these CMDs indicates that the SFH solutions shown in Figures~\ref{SFRbright} and~\ref{SFHbright} constitute an overall good fit of the data, with CMDs largely consistent within 3 sigma in most bins for both the solutions with and without taking into the MDF. The synthetic CMDs correctly reproduce all evolutionary features in the lower CMD, including the location and spread of the MSTO, SGB and faint RGB as well as the presence of younger populations above the nominal SGB. However, Figure~\ref{Hessbright} also shows features that do not fit as well, such as the presence of red RGB stars which could be fit to residual foreground stars and stars red-ward of the faint MS. 

\begin{figure}
\centering
\includegraphics[angle=0, width=0.49\textwidth]{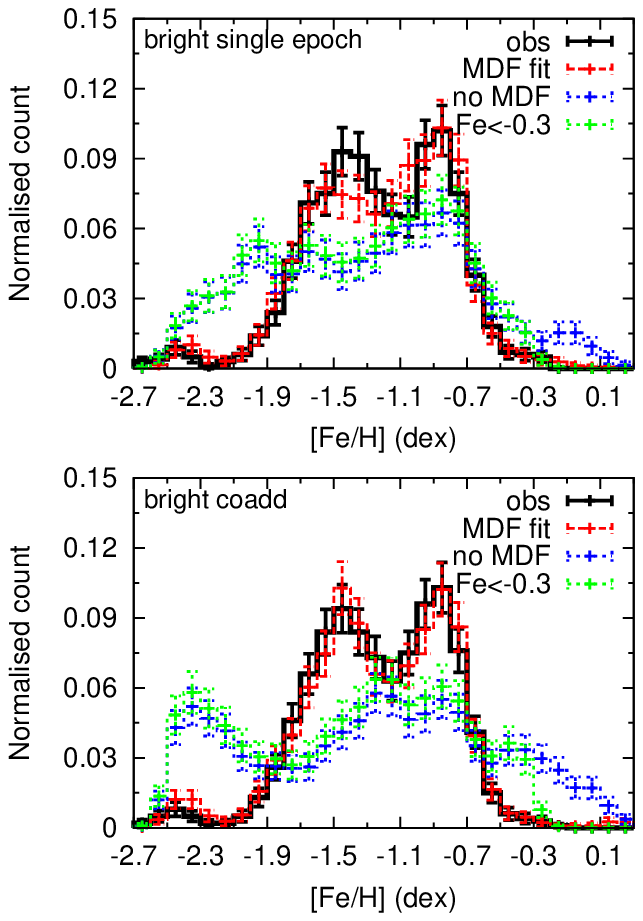}
\caption{Normalised metallicity distribution function~(as sampled on the bright RGB) for the Sgr bright stream from deep co-added~(top) and single-epoch~(bottom) photometry, with and without MDF fitting in the SFH derivation. The observed MDF is shown as black lines, while the best-fit result from the SFH derivation is shown in red and blue for solution respectively with and without MDF fitting. The green line indicates the MDF derived without MDF fitting after applying a cut on the possible metallicity of stream member stars. The error bars indicate the uncertainties on the MDF due to uncertainties on the SFH for the model and due to Poisson errors for the observations. \label{MDFbright}}
\end{figure}

Comparison between the derived synthetic MDF for each solution~(see Figure~\ref{MDFbright}) and the observations shows a good fit for the solution with MDF fitting, which is expected since the MDF was used directly as an input. For the solution without MDF fitting, the best-fit model is broadly consistent with the observations, with the mean peak at similar metallicities. However, the distribution of more metal-poor stars is much broader than the observations, extending all the way to [Fe/H]=$-$2.3 instead of peaking at [Fe/H]=$-$1.5. The origin of this excess of metal-poor stars in the model is not clear, but could be related to the similarity between metal-poor models in the CMD at the MSTO level, leading to a smearing out of metallicity information. Furthermore, it is also possible that these stars are present in the Sgr stream, but not fully sampled in the observed MDF. A slight excess of metal-rich stars is also visible, which might be linked to stellar populations fit to the red foreground stars seen in Figure~\ref{Hessbright}. 

The best-fit SFH solution for the Sgr bright stream is characterised by a well-defined sequence in age-metallicity space~(see Figure~\ref{SFRbright}) starting from old, metal-poor populations~([Fe/H]$\approx$$-$2.5, age$\approx$14 Gyr) and extending to intermediate ages~($\approx$5 Gyr) with high metallicities~([Fe/H]=$-$0.7). This sequence is relatively narrow, indicating that Sgr experienced metal enrichment under relatively isolated conditions without strong inhomogeneous mixing. The sequence of Sgr populations displays a change of slope in age-metallicity space at an age$\approx$12-13 Gyr and metallicity [Fe/H]$\approx$$-$1.5. The location of this change in slope of the age-metallicity relation~(AMR) is roughly consistent with that of the $\alpha$-element knee observed by~\citet{deBoer2014A}, indicating it could be linked to the onset of supernovae type Ia contributing to the metal enrichment of Sgr. 

To compare the results obtained for the Sgr stream in Stripe~82 to
results for Sgr across a wider spatial range, we compare our SFH to
the stellar population parameters of 11 globular clusters~(GCs)
associated to the Sgr stream, as determined from very deep CMD
studies~\citep[][and references therein]{Forbes10}. The age and
metallicity~(and uncertainties) are shown as large circles with error
bars overlaid on the Sgr stream results in Figure~\ref{SFRbright}. The
sequence traced out by Sgr populations is consistent with the
parameters determined for the GCs indicating that the SFH obtained
here is indeed a good representation of Sgr populations.

Results obtained with and without the spectroscopic MDF fitting show
an overall good agreement in recovering the same sequence in
age-metallicity space. However, the SFH results without including the
MDF show the presence of very metal-rich
populations~([Fe/H]$\ge$$-$0.3), which are not recovered in the
spectroscopic metallicities of stream stars at this location on the sky. The red reference isochrone in Figure~\ref{SgrbrightCMD}
traces the location in the CMD of the metal-rich populations seen in
Figure~\ref{SFRbright} and shows these are inconsistent with the
observed upper RGB of Sgr if given a significant SFR. Therefore, these
populations are likely the result of fitting to red excess foreground
stars and do not represent genuine Sgr populations. To assess the effect of these population on the synthetic MDF, the green, dotted line in Figure~\ref{MDFbright} shows the synthetic MDF as obtained without including stars with metallicities above [Fe/H]=$-$0.3. This MDF provides a better fit to the observed MDF at high metallicities, while metal-poor stars are still spread over a wide range of metallicities. 

The SFH as a function of age and metallicity shows that Sgr displays
high SFR across a large range of ages, with several periods of increasing and decreasing SFR.
Below an age of
$\approx$6 Gyr the SFR declines rapidly, indicating the end of star
formation in the part of Sgr associated to the stream at this
location. This might be related to the accretion of Sgr by the MW,
which is expected to strip the dwarf galaxy of its remaining gas and
truncate star formation. The stellar populations in the Sgr stream
span a range of metallicity between [Fe/H]$\approx$$-$2.5 and
[Fe/H]=$-$0.5 without a strong peak at metal-poor metallicities seen
in many other dwarf
galaxies~\citep[e.g.,][]{deBoer2012A,deBoer2012B}. This lack of a
dominant metal-poor component might be related to the large mass of
Sgr, resulting in efficient metal enrichment at early times or this
section of stream being stripped from the more central parts of the
progenitor, dominated by more metal-rich stars.

The distinct periods of star formation in Figures~\ref{SFRbright}
and~\ref{SFHbright} as constrained using stars from the bottom of the CMD 
are in good agreement with evolved population features in the bright CMD~(RGB, HB and RC).
The position and slope of the RGB as traced by the
isochrones shown in Figure~\ref{SgrbrightCMD} agrees with the 
spectroscopic metallicities shown in Figure~\ref{SgrbrightCMD}.
This indicates that these populations provide a good fit to both evolved
and non-evolved CMD indicators. 
Results obtained with and without MDF
fitting are consistent within the errors, except for the presence of
very metal-rich populations in the solution without spectroscopic MDF.

\subsection{Deep co-added data}

Due to the better quality of the deep SDSS co-added data, we achieve a
higher S/N of the Sgr stream and smaller photometric errors at the
level of the nominal MS turnoff~(see Figure~\ref{SgrbrightCMD}). This
allows us to determine the effects on the SFH related to the depth and
data quality of SDSS photometry.

The CMD residuals in the right panels of Figure~\ref{Hessbright} once
again show an overall good fit of the synthetic CMDs for both the
solutions with and without taking into account the MDF. However, a
strong positive residual is present at g-i$\approx$0.3, which is
consistent with the blue edge of the CMD area affected by MW
foreground contamination. Therefore, this residual could be the result
of a different population mix between the stream and foreground
regions, which do not sample identical Galactic latitude and longitude
within Stripe~82. Furthermore, the sharp blue edge of the residual
could also indicate a difference in extinction not accounted for in
dust maps or different photometric error distribution. Furthermore,
the CMD residuals also show the presence of several populations
clearly inconsistent with the main locus of Sgr stars, such as fits to
very red foreground stars or blue BSS stars. These populations show up
as low level SFR in Figure~\ref{SFRbright} as
metal-rich~([Fe/H]$<$$-$0.3), old populations or stars with very
young~($<$2 Gyr) ages. 

The SFH as obtained from the deep co-added data is overall in good
agreement with the SFH obtained from single-epoch data. The
well-defined sequence in age-metallicity space is also visible in the
co-added data and consistent with the shallower data. The sequence
appears more defined and narrower in the deep photometry, due to the
better data quality, resulting in better age resolution. Comparison
between the deep and shallow data shows that the single-epoch SDSS
data quality is sufficient to correctly recover the age-metallicity
relation of the Sgr stream, giving further confidence to the results
of Section~\ref{brightsingleepoch}.

\begin{figure*}
\centering
\includegraphics[angle=0, width=0.9\textwidth]{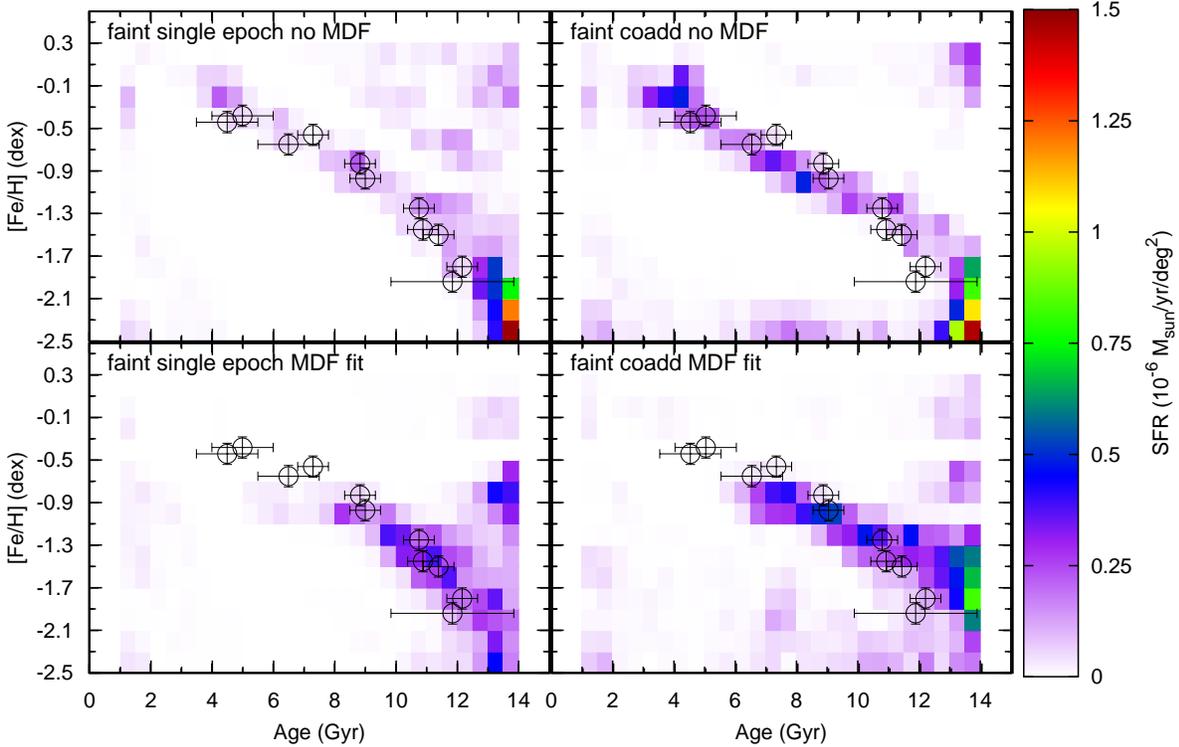}
\caption{Full SFH solution as a function of age and metallicity for the faint Sgr stream within Stripe 82~(see Figure~\ref{Sgrspatial}), with and without MDF fitting in the SFH derivation. Left panels show the solution derived from single-epoch photometric data, while the right panels shows solution derived from deep co-added data. \label{SFRfaint}}
\end{figure*}

The SFH projected on age and metallicity also shows similar features
as seen for single-epoch data in Figure~\ref{SFHbright}, with strong
star formation from ancient times~($>$10 Gyr) and a strong decline in
SFR at $\approx$6 Gyr ago. Results obtained with the inclusion of MDF
fitting for both single-epoch and co-added data show peaks at similar
times in Figure~\ref{SFHbright} consistent within the uncertainties,
although results from co-added photometry are shifted slightly to
older ages. The co-added SFH shows a drop in SFR around an age of 9
Gyr, which is not reproduced in the top panels of
Figure~\ref{SFHbright}. The most likely explanation for this is the difference in data
quality between both datasets, resulting in a worse age resolution in
the single-epoch data and smoothing out the SFR peaks to fill the gap
at 9 Gyr. Furthermore, this feature might also be related to the
broad SGB in Figure~\ref{Hessbright}, showing gaps in the density
distribution. Unfortunately, this region of the CMD suffers from poor
MW foreground decontamination, which may affect the SFH results.

\begin{figure*}
\centering
\includegraphics[angle=0, width=0.9\textwidth]{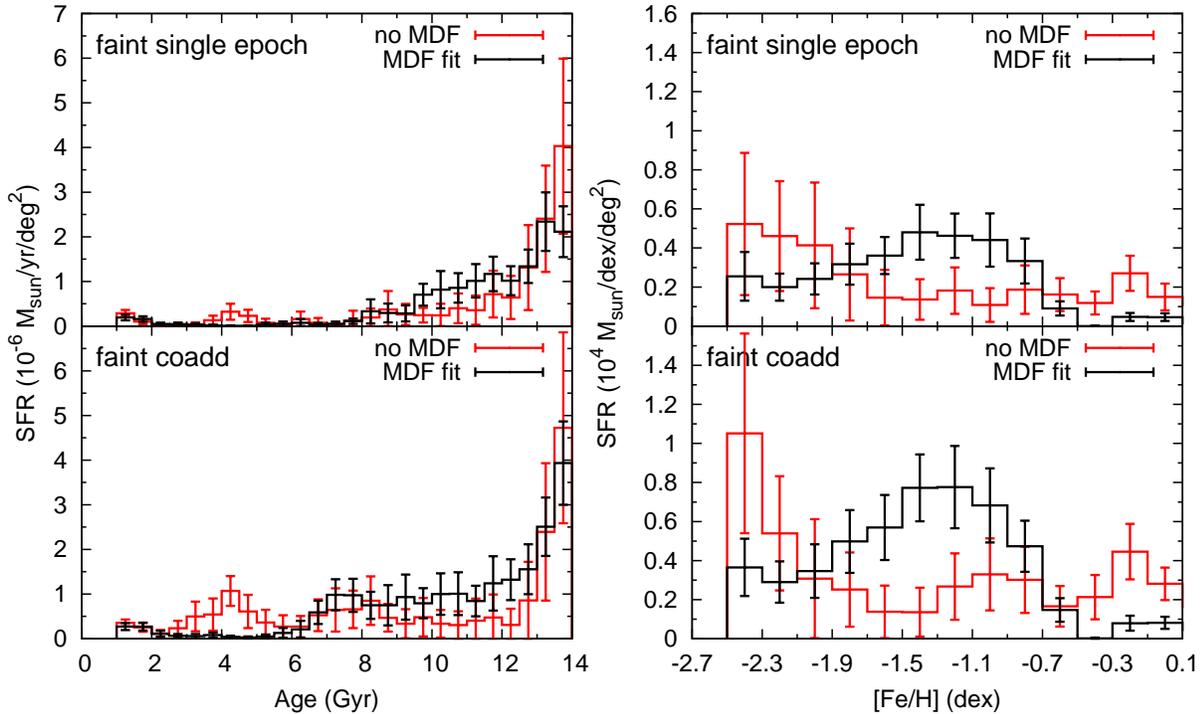}
\caption{Star formation rates of the faint stream projected on age~(star formation history, left) and metallicity~(chemical evolution history, right). Solutions from single-epoch data are shown in the top panels, while co-add photometry solutions are shown in the bottom panels.  \label{SFHfaint}}
\end{figure*}

\section{Star formation history of the faint stream}
\label{faintstream}

To study the properties of the faint Sgr stream, we employ the same
procedure as used in Section~\ref{brightstream}. To compare the
results of the faint stream to those of the bright stream, we use the
same range of stellar populations in the fitting with the same age and
metallicity resolution. The faint stream SFH is fit using the (i,
g$-$i) CMD, with the same CMD limits as assumed in the fitting of the
bright stream. However, due to the lower number density of stars in
the faint stream, the colour and magnitude bin size has been doubled
to achieve enough signal in each CMD bin. The fitting routine once
again makes use of completeness information based on the comparison
between deep co-added and single-epoch photometry as shown in
Figure~\ref{SgrfaintCMD}.
\begin{figure*}
\centering
\includegraphics[angle=0, width=0.45\textwidth]{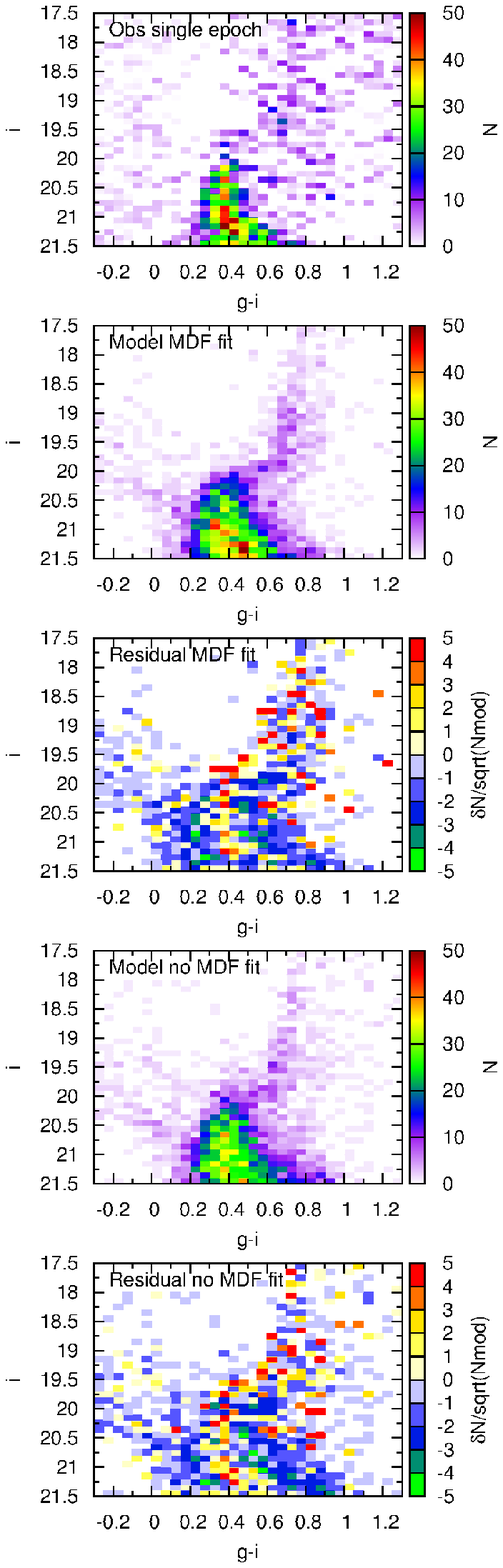}
\includegraphics[angle=0, width=0.45\textwidth]{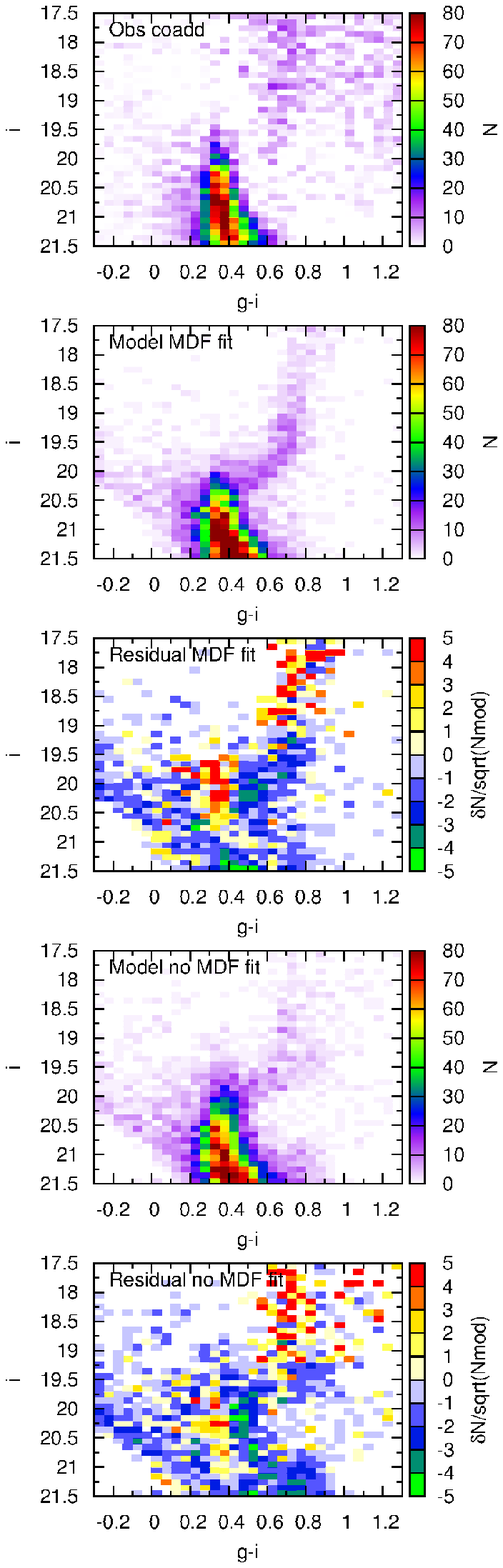}
\caption{Observed and best-fitting CMDs for the faint Sgr stream from single-epoch~(left) and deep co-added~(right) photometry, with and without MDF fitting. The middle and bottom rows show the difference between the observed and best-fit CMDs, expressed as a function of the uncertainty in each CMD bin.  \label{Hessfaint}}
\end{figure*}

\subsection{Single-epoch data}
\label{faintsingleepoch}

The best-fit SFH solution for the faint Sgr stream is shown in
Figure~\ref{SFRfaint} as a function of age and
metallicity. Figure~\ref{SFRfaint} also shows the solution as obtained
without fitting the spectroscopic MDF. The SFR values are normalised
by the spatial area of each region of the stream, using a spatial area
of 24 deg$^{2}$ for the faint stream in Stripe~82. Furthermore,
Figure~\ref{SFHfaint} shows the results for the SFH and CEH by
projecting the SFR onto age and metallicity respectively, in units of
solar mass per year and dex respectively. 

\begin{figure}
\centering
\includegraphics[angle=0, width=0.49\textwidth]{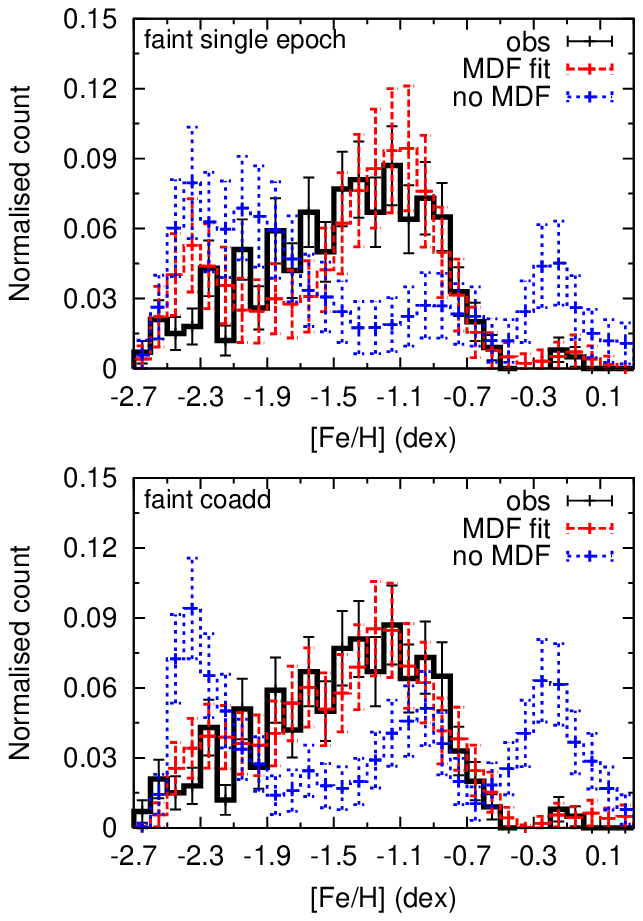}
\caption{Normalised metallicity distribution function for the Sgr faint stream, with and without MDF fitting in the SFH derivation. Deep co-added solutions are shown in the top panel and single-epoch solutions in the bottom panel. The observed MDF is shown as black lines, while the best-fit result from the SFH derivation is shown in red and blue for solution respectively with and without MDF fitting. Error bars indicate the uncertainties on the MDF due to uncertainties on the SFH for the model and due to Poissonian errors for the observations.\label{MDFfaint}}
\end{figure}

Comparisons between the observed (i, g$-$i) CMD and the best-fit CMD
obtained from the SFH are shown in Figure~\ref{Hessfaint} for each
region, along with the residual in each bin in terms of Poisson
uncertainties. To further asses the quality of the fits,
Figure~\ref{MDFfaint} shows the observed MDFs on the upper RGB in comparison to the
synthetic MDFs inferred from the SFH solution with and without MDF
fitting. Once again, for the solution including spectroscopic MDF
fitting, the good fit is expected given that the MDF is used as
input. 

The CMD residuals indicate that the SFH solution produces an overall
good fit of the observed CMDs, without any obvious systematic
residuals. The synthetic CMDs reproduce the MS and MSTO at the right
location and produce an SGB and lower RGB consistent with the observed
CMD as sampled from both photometric and spectroscopic stream
members~(see Figure~\ref{SgrfaintCMD}). Due to the lower intrinsic
signal of the faint stream, the sampling of the stellar evolution
features is less than for the bright stream, leading to greater
uncertainties in the SFH. Furthermore, there is a stronger effect of
incomplete MW foreground correction, especially visible red-ward of
the nominal MSTO and SGB~(g$-$i$\approx$0.8,
i$\approx$20.5). Therefore, Figure~\ref{SFRfaint} shows the presence
of more anomalous populations with higher metallicities and older
ages~(driving isochrones to redder colours) compared to the Sgr
stellar population sequence. 

The SFH results presented in Figures~\ref{SFRfaint} and~\ref{SFHfaint}
show that the Sgr faint stream displays a sequence in age-metallicity
space similar to that observed in the bright stream~(see
Figure~\ref{SFRbright}). This sequence is also compatible with the
age-metallicity relation traced out by GCs associated to the Sgr dwarf
galaxy, although the faint stream does not sample the same extent of
the sequence as the bright stream. Instead, the faint stream samples
the older, more metal-poor part of the sequence, being comprised
mostly of stars with [Fe/H]$<$$-$1 and age$>$9 Gyr. This is also
visible in Figure~\ref{SFHfaint}, showing that star formation peaks at
ancient ages.

The SFH results obtained with and without MDF fitting both trace the
same age-metallicity sequence. However, the SFH peak in the solution
without MDF fitting is strongly peaked at metal-poor metallicities and
is more spread out in age-metallicity space than the solution with MDF
fitting. Comparison to Figure~\ref{Hessfaint} shows that the solution
without MDF indeed displays a dominant blue SGB and RGB consistent
with metal-poor, old populations, while the MDF fit produces a redder
RGB and SGB due to more metal-rich populations. Furthermore, best-fit
models without MDF result in a wide range in colour at faint
magnitudes, with fits to stars associated to MW halo
contamination. Due to the sparse sampling of the evolved features in
the CMD and the strong MW contamination it is unclear which of these
solutions fits the CMD best. Deeper CMDs or wider angle along the
stream are needed to increase the S/N of the stream signal to
unambiguously determine the presence of these features.

Comparing the results from Figures~\ref{SFRbright} and~\ref{SFRfaint}
shows that both the bright and faint stream in Stripe~82 are drawn
from the same stellar population mix present elsewhere in the stream
and in the progenitor. However, the populations in the faint stream
are not consistent with those of the bright stream, arguing 
in favour of a scenario where the faint stream consists of material stripped from the outskirts of the Sgr dwarf at an earlier epoch.

\subsection{Deep co-added data}
\label{faintcoadd}

To obtain a better picture of the faint stream stellar populations, we
once again make use of the deep co-added photometry of Stripe~82. Due
to the deeper data we achieve a higher S/N not only for the MS region
of the CMD but also for the MSTO and SGB regions, crucial for
disentangling the ages of the stellar populations.

Figures~\ref{SFRfaint} and~\ref{SFHfaint} present the best-fit SFH
solution for the faint Sgr stream from co-added photometry. The SFR as
a function of age and metallicity is shown in Figure~\ref{SFRfaint} in
comparison to single-epoch results, while Figure~\ref{SFHfaint}
displays the SFR projected onto the age and metallicity axes
respectively. The comparison between observed and synthetic CMD is
shown in the right panels of Figure~\ref{Hessfaint} for the solution
with and without MDF fitting, along with associated
residuals. Additionally, the comparison between observed and best-fit
MDFs is shown in Figure~\ref{MDFfaint}.

Comparison between the single-epoch and co-added CMDs in
Figures~\ref{Hessfaint} shows that the MW foreground contamination is
greatly reduced around the MSTO/SGB region, leading to a better
determination of the ages of the faint stream populations. Similar to
results obtained for the bright stream, systematic residuals are
visible in Figure~\ref{Hessfaint} above the nominal MSTO and at the
faint edge of the CMD, for both the solutions with and without taking
into the MDF. The location of the positive residuals coincides with a
region of the CMD strongly affected by MW foreground correction. Due
to the limited choice of foreground region available within the deep
co-added data, it is likely that the MW stellar population in between
the foreground and stream regions is not fully consistent, which can
account for the residuals in Figure~\ref{Hessfaint}.

The SFH of the faint stream as obtained from the co-added data is
consistent with the single-epoch SFH shown in Figures~\ref{SFRfaint}
and~\ref{SFHfaint}. Similar sequences are visible in both solutions,
with peaks at similar age and metallicity. Due to the better data
quality of the deep data, the sequence is more defined in the co-added
data and shows less effects of populations fit to residual foreground
stars. Results for the SFH projected on age and metallicity as well as
the spectroscopic MDF are also consistent to within the error
bars. The SFH is dominated by old star formation at relatively
metal-poor metallicities~([Fe/H]$<$$-$1.5) but also shows a peak at
age $\approx$8 Gyr ago, slightly older than single-epoch results.

The synthetic MDF derived from the best-fit SFH solution is in better
agreement with the observed MDF given the generous error bars, unlike
the results for single-epoch data. This metal-rich peak is related to
the burst of star formation $\approx$8 Gyr ago, similar to what was
observed for the bright stream in Figure~\ref{SFRbright}.

\section{Conclusions and discussion}
\label{conclusions}

In this work, we present the first detailed quantitative study of the
stellar populations of the Sgr trailing stream, using photometric and
spectroscopic observations from the SDSS surveys. By modelling the
available data, we infer the Sgr SFH and MDF as displayed by the stars
in a portion of the (bifurcated) trailing tail of the galaxy. Such
detailed SFH and MDF of the bright and faint stream components within
the Stripe~82 region provide new constraints of the model of the dwarf
disruption. The results of our study can be summarised as follows.

\begin{itemize}

\item As displayed in Figures~\ref{Hessbright} and ~\ref{Hessfaint},
  the models of the CMD density distribution appear sensible and
  reproduce the bulk of the Sgr stellar populations as observed by the
  SDSS in the Southern Hemisphere. There is, however, a small amount
  of positive residuals, in particular at the MSTO colour. This
  systematic mismatch is possibly due to a small level of asymmetry in
  the Galactic disk and halo, and therefore was introduced at the
  foreground subtraction stage. Alternatively, it can not be ruled out
  that a slightly misjudged extinction could cause such an artificial
  pile-up.

\item Solutions with MDF and without MDF look generally consistent,
  however fits without MDF constraints are clearly degenerate and appear
  to lead to biased inference as to the SFH details (see Figures
  \ref{SFRbright}-\ref{MDFfaint}). For example, solutions derived without including MDF fitting
  produce a more extended sequence reaching all the way up to solar
  metallicity, which is not supported by the spectroscopic
  observations. Comparison with the observed CMD~(see red isochrone in
  Figure~\ref{SgrbrightCMD}) shows that populations with these
  parameters are not observed in great number on the red side of the
  RGB locus or in the sample of kinematic member stars, indicating
  that these populations should not be present with substantial star
  formation rates. The presence of these populations could be a result
  of fitting models to residual foreground stars red-ward of the
  nominal RGB. However, the age and metallicity of the populations is
  consistent with a picture of steady chemical enrichment in an
  isolated environment and does occur in the Sgr dwarf galaxy as
  traced by the distribution of GCs. To improve the interpretation of
  these populations, we need to sample a larger portion of the
  streams, to increase the relative S/N of these stream populations.

Additionally, comparison of results derived with and without MDF
fitting shows that solutions without MDF fitting produce a
systematically higher SFR in populations with [Fe/H]$>$$-$2. This is
due to the lack of stars with such low metallicities in the
spectroscopic MDF, as shown in Figures~\ref{MDFbright}
and~\ref{MDFfaint}. This lack of metal-poor stars could be a result of
the efficient enrichment in a massive dwarf galaxy such as Sgr,
quickly increasing the metallicity of the interstellar medium from
which subsequent generations of stars are formed. On the other hand,
the region of the CMD occupied by metal-poor stars is heavily
contaminated by foreground stars, leading to a larger fraction of
fibers being assigned to foreground stars than for metal-rich stars,in
the case of random spectroscopic fiber assignment. This type of
completeness cannot be corrected for given the inherently unknown
distribution of foreground and background populations. A better
spectroscopic sampling of the Sgr stream is necessary to unambiguously
determine the fraction of metal-poor stars within the Sgr stream.

\item Both bright and faint stream components show a tight sequence in
  the plane of Age vs [Fe/H] as shown in Figures~\ref{SFRbright}
  and~\ref{SFRfaint} indicating that star formation within Sgr took
  place in a well-mixed medium, homogeneously enriched in metals over
  $\approx$8 Gyr.  The tight sequence starts from old, metal-poor
  populations and extends to a metallicity of [Fe/H]$\approx$$-$0.7 at
  an age of $\approx$5 Gyr before star formation terminates.  The SFH
  (see Figures~\ref{SFHbright} and \ref{SFHfaint}) confirms the
  extended formation history of the progenitor Sgr dwarf galaxy and
  quantifies the strength and stellar population make-up of each star
  formation episode. The tight sequence observed in the bright stream
  is also reproduced in the faint one, although the lower S/N of the
  stream results in the presence of more anomalous populations due to
  residual foreground stars. Star formation in the faint stream is
  dominated by metal-poor populations, with a SFH mostly composed of
  stars $>$10 Gyr old~(see Figure~\ref{SFHfaint}). Therefore, the
  faint stream appears composed of a simpler stellar population mix
  than the bright stream.

\item To compare the SFH of the streams to the overall SFH of Sgr,
  Figures~\ref{SFRbright} and~\ref{SFRfaint} show the age and
  metallicity of GCs associated to the Sgr dwarf
  galaxy~\citep{Forbes10}. The GCs trace out the same tight sequence
  in age and metallicity space, indicating that the sequence observed
  in both streams is consistent with Sgr populations present elsewhere
  in the stream and main body. Therefore, both streams are consistent
  with being drawn from the mix of populations associated to Sgr.

\item We can use the SFH derived here for both stream components to
  study the formation history of the parent Sgr dwarf galaxy. It is
  clear from Figure~\ref{SFHbright} that the Sgr galaxy has undergone
  an extended formation history, with multiple peaks in SFR. This indicates that Sgr formed stars
  over a substantial period of star formation of at least 7 Gyr. The
  MDF of the bright stream also shows strong evidence for a
  bi-modality with peaks at [Fe/H]$\approx$-1.5 and -1. 

The faint stream, on the other hand, displays neither clear peaks in
the SFH nor additional bumps in the MDF. We conclude again (bearing in
mind that the faint stream data possesses lower S/N), that the fainter
stream has experienced a less eventful star formation history and
contains a simpler population mix. 

\item In the bright stream, star formation rates drop rapidly around
  $\approx$5-7 Gyr ago; this also corresponds to the last substantial
  star formation activity in the faint stream. This shutdown of star
  formation could be caused by the infall of Sgr into the MW
  potential, coinciding with stripping of gas from the outskirts of
  the system, from which the streams were formed.

\item The sequence of Sgr populations displays a change of slope in
  age-metallicity space at an age between 11-13 Gyr and metallicity
  [Fe/H]$\approx$$-$1.5. The location of this change in slope of the
  age-metallicity relation~(AMR) is consistent with that of the
  $\alpha$-element knee observed by~\citet{deBoer2014A}, indicating that
  supernovae type Ia started contributing noticeably to the abundance
  pattern $\approx$1-3 Gyr after the start of star formation in Sgr.

\item There is one additional significant difference between the
  stellar population properties of the bright and the faint
  streams. As illustrated in the Figure~\ref{MDFobs}, the faint stream seems to
  lack a significant metal-rich component with [Fe/H]$>$-0.9, the
  population easily discernible in the bright stream.

\item Finally, we would like to point out that the analysis has been
  carried out using both single-epoch and stacked SDSS photometry. It
  is re-assuring to see that the results of modelling of each dataset
  are consistent with each other. However, deeper data clearly
  provides better constraints on the SFH: the age-metallicity sequence
  is much tighter, while the peaks corresponding to SFH activity are
  more significant.

\end{itemize}


Overall, we believe that this pilot study has convincingly
demonstrated that SFH and MDF inference can be obtained for
low-surface brightness structures in the Galactic halo. 
Studies such as this one will greatly benefit from future deep wide-field surveys such as LSST, which will be able to tap into the large number of MS stars in both stream components, leading to more accurate inferred distances and stellar population parameters. Apart from the obvious necessity for a deeper wide-area survey data, we would also like to emphasise two potential routes for improvement. First, to avoid biases
caused by improper foreground subtraction, analysis of the broad band
stream photometry and spectroscopy could proceed simultaneously with
the foreground modelling. For the Galaxy, this of course would require
simultaneous fitting of the volume density as well as the CMD. Second,
small but noticeable CMD density residuals displayed above could
potentially be a sign of changing stream distance, distance spread or the presence of
additional stream components. Therefore, rather than fixing the
sub-structure's distance and distance spread, as well as the number of unrelaxed fragments
along the line-of-sight, the model can be relaxed to include these as
free parameters.

The results of our analysis bear implications for the modelling of the
Sgr disruption. Interestingly, as summarised earlier, the star
formation activity, inferred with the stream stars, appears curtailed
$\approx$ 5-7 Gyr ago. Typically, in stellar stream modelling, the total
disruption time is the least constrained ``nuisance'' parameter which
can play, however, a very important role. For example, in case of the
Sgr dwarf, if the tidal debris have orbited within the MW for
sufficiently long time, the global stream parameters, such as apsidal
and orbital plane precession angles could be affected by the
subsequent material accretion. As far as the differences between the
bright and the faint stream components are concerned, we i) prove that
the Sgr dwarf is likely the progenitor of the faint component as well
as the bright one, and ii) present strong evidence that the
star-formation and chemical enrichment proceeded in a similar fashion
for each part of the bifurcation. However, we also confirm earlier
claims of \citet{Koposov12} of a subtle variation in the MDF of the
faint stream as compared to the bright. According to our analysis, the
faint component of the trailing stream displays a simpler mix of
stellar populations dominated by old metal poor stars and lacks a
strong metal-rich ingredient obviously present in the bright
tail. Viewed naively, such differences in stellar populations can
perhaps be explained if the faint stream was produced by the material
stripped i) earlier and ii) from the outskirts of the dwarf.

\section*{Acknowledgements}
The research leading to these results has received funding from the
European Research Council under the European UnionÕs Seventh Framework
Programme (FP/2007-2013) / ERC Grant Agreement n. 308024. VB
acknowledges financial support from the Royal
Society. T.d.B. acknowledges financial support from the ERC. The authors would also like to thank the referee for his or her comments.

Funding for SDSS-III has been provided by the Alfred P. Sloan
Foundation, the Participating Institutions, the National Science
Foundation, and the U.S. Department of Energy Office of Science. The
SDSS-III web site is http://www.sdss3.org/.

SDSS-III is managed by the Astrophysical Research Consortium for the
Participating Institutions of the SDSS-III Collaboration including the
University of Arizona, the Brazilian Participation Group, Brookhaven
National Laboratory, Carnegie Mellon University, University of
Florida, the French Participation Group, the German Participation
Group, Harvard University, the Instituto de Astrofisica de Canarias,
the Michigan State/Notre Dame/JINA Participation Group, Johns Hopkins
University, Lawrence Berkeley National Laboratory, Max Planck
Institute for Astrophysics, Max Planck Institute for Extraterrestrial
Physics, New Mexico State University, New York University, Ohio State
University, Pennsylvania State University, University of Portsmouth,
Princeton University, the Spanish Participation Group, University of
Tokyo, University of Utah, Vanderbilt University, University of
Virginia, University of Washington, and Yale University.
\bibliographystyle{mn2e_fixed}
\bibliography{Bibliography}

\label{lastpage}

\end{document}